\documentclass[10pt,twocolumn,letterpaper]{article}

\usepackage{iccv}
\usepackage{times}
\usepackage{epsfig}
\usepackage{graphicx}
\usepackage{amsmath}
\usepackage{amssymb}
\usepackage{svg}
\usepackage{multirow}
\usepackage{authblk}
\usepackage[absolute,overlay]{textpos}
\usepackage{todonotes}
\usepackage{booktabs}
\usepackage{tabularx}
\usepackage{pifont}
\usepackage[pagebackref=true,
breaklinks=true,letterpaper=true,colorlinks,bookmarks=false]{hyperref}
\usepackage[capitalize]{cleveref}

\usepackage[pagebackref=true,breaklinks=true,letterpaper=true,colorlinks,bookmarks=false]{hyperref}

\iccvfinalcopy 

\def\iccvPaperID{CVAMD-67} 

\ificcvfinal\fi

\newcommand{\email}[1]{\texttt{\small #1}}
\begin{document}

\begin{textblock*}{\textwidth}(18mm,12mm)
    \vspace{2mm}
    \tiny
    \centering
    Ethan Dack, Lorenzo Brigato, Matthew McMurray, Matthias Fontanellaz, Thomas Frauenfelder, Hanno Hoppe, Aristomenis Exadaktylos, Thomas Geiser, Manuela Funke-Chambour, Andreas Christe, Lukas Ebner, and Stavroula Mougiakakou.\\
    ``An Empirical Analysis for Zero-Shot Multi-Label Classification on COVID-19 CT Scans and Uncurated Reports.''\\
    \textit{Proceedings of the IEEE/CVF International Conference on Computer Vision (ICCV) Workshops 2023.}\\
    \copyright\ 2023 IEEE. Personal use of this material is permitted.
    Permission from IEEE must be obtained for all other uses, in any current or future media, including reprinting/republishing this material for advertising or promotional purposes, creating new collective works, for resale or redistribution to servers or lists, or reuse of any copyrighted component of this work in other works.
    The final publication will be available at
    \href{https://ieeexplore.ieee.org/}{ieeexplore.ieee.org}.\\
    \vspace{2mm}
\end{textblock*}

\title{An Empirical Analysis for Zero-Shot Multi-Label Classification on COVID-19 CT Scans and Uncurated Reports}

\author[1,2]{Ethan Dack}
\author[1,2]{Lorenzo Brigato}
\author[3]{Matthew McMurray}
\author[1,2]{\authorcr Matthias Fontanellaz}
\author[3]{Thomas Frauenfelder}
\author[2,5,6]{Hanno Hoppe}
\author[7]{Aristomenis Exadaktylos}
\author[8]{Thomas Geiser}
\author[8]{Manuela Funke-Chambour}
\author[3]{Andreas Christe}
\author[3]{Lukas Ebner}
\author[1,2]{Stavroula Mougiakakou}

\affil[1]{AI in Health and Nutrition, ARTORG Center for Biomedical Engineering Research}
\affil[2]{University of Bern}
\affil[3]{Department of Diagnostic, Interventional, and Pediatric Radiology, Bern University Hospital}
\affil[4]{Diagnostic and interventional Radiology, University Hospital Zurich}
\affil[5]{Department of Radiology, Lindenhofspital, Bern}
\affil[6]{Campus Stiftung Lindenhof Bern (SLB), Bern, Switzerland}
\affil[7]{Department of Emergency Medicine, Bern University Hospital}
\affil[8]{Department of Pulmonary Medicine, Bern University Hospital}
\affil[ ]{\email{\{ethan.dack,lorenzo.brigato,matthias.fontanellaz,
hanno.hoppe,stavroula.mougiakakou\}@unibe.ch\\
\{matthewthomas.mcmurray,aristomenis.exadaktylos,thomas.geiser,
manuela.funke-chambour,andreas.christe,lukas.ebner\}@insel.ch\\
thomas.frauenfelder@usz.ch}}

\maketitle
\ificcvfinal\thispagestyle{empty}\fi

\begin{abstract}
The pandemic resulted in vast repositories of unstructured data, including radiology reports, due to increased medical examinations.
Previous research on automated diagnosis of COVID-19 primarily focuses on X-ray images, despite their lower precision compared to computed tomography (CT) scans.
In this work, we leverage unstructured data from a hospital and harness the fine-grained details offered by CT scans to perform zero-shot multi-label classification based on contrastive visual language learning.
In collaboration with human experts, we investigate the effectiveness of multiple zero-shot models that aid radiologists in detecting pulmonary embolisms and identifying intricate lung details like ground glass opacities and consolidations.
Our empirical analysis provides an overview of the possible solutions to target such fine-grained tasks, so far overlooked in the medical multimodal pretraining literature.
Our investigation promises future advancements in the medical image analysis community by addressing some challenges associated with unstructured data and fine-grained multi-label classification.
\end{abstract}

\section{Introduction}

Artificial intelligence (AI) research in the medical domain throughout the pandemic prioritized applying supervised learning methods to help hospitals diagnose or triage patients faster. 
Whilst models were developed at a fast pace, the majority of these were deemed not fit for clinical use \cite{covid-review1}.
Despite these setbacks, the pandemic led to the generation and collection of substantial medical imaging data, among others.
AI is still considered a high-quality strategy to assist in diagnosis, severity assessment, and prognosis of long COVID-19 \cite{covid-review2}.
The large amounts of unlabelled data from the pandemic allow the possibility of exploring self-supervised learning models to be developed.
One of the most popular methods is contrastive visual language pertaining (CLIP), which enables training on pairs of images and text with zero-shot capabilities \cite{CLIP}.
CLIP eliminates the need for precisely annotated datasets and learns representation from noisy image-text pairs, potentially resulting in significant time and cost savings.

Applying self-supervised deep learning methods to open-source data has rapidly gained popularity in medical and non-medical domains \cite{SimCLR, self-supervised-review, Chexzero, Convirt}.
The current success of self-supervised learning, particularly contrastive, can be attributed to preventing dimension collapse through aggressive data augmentation and negative pairs \cite{Lijing} and utilizing large datasets.
For instance, CLIP training from scratch was enabled by access to 400 million pairs of images and texts crawled from the web.
Contrastive methods have been highly valuable in downstream tasks like image classification, enabling the creation of competitive representations comparable to fully supervised networks \cite{CVPR2021, CVPR2022}.

The medical domain does not have access to such large datasets since collecting data is costly and requires highly specialized human expertise \cite{wang2017chestx, wang2020covid}.
Despite this, there has recently been a successful application of multimodal contrastive pretraining techniques \cite{Convirt, Chexzero, biomedclip, medclip}.
Most of the aforementioned studies performed pretraining on X-rays \cite{Convirt, Chexzero}, which are easier to gather but less precise than other modalities, e.g., CT scans. 
Furthermore, given the smaller scale of biomedical datasets and the significant domain shift among different subdomains compared to natural images (e.g., X-rays to CT scans), it remains an open research question on how to adapt and fine-tune available pre-trained models properly \cite{medclip}.

This work focuses on fine-tuning pre-trained encoders via CLIP on CT images and uncurated radiology reports obtained during the pandemic.
We investigate several issues deriving from fine-tuning on a different domain, such as the data preprocessing of large volumetric CT scans and long unstructured reports.
Furthermore, in collaboration with expert radiologists, we establish a fine-grained multi-label classification task that evaluates disease severity and identifies the presence of five distinct characteristics commonly associated with COVID-19:  pulmonary embolism, pneumonia, consolidation, infiltrates and ground glass opacities.
The zero-shot classification task is particularly challenging due to the uncurated structure of the reports and the fine-grained nature of the task.
To improve the correct matching across visual predictions and text targets, rather than keeping class-independent templates like standard practice \cite{CLIP}, we design per-class templates.  

We specifically focused on patients diagnosed with COVID-19, as we aim to develop a valuable tool to aid radiologists in identifying individuals at the highest risk. 
More broadly, we hope that our empirical analysis could be helpful for researchers involved in deploying pre-trained models on different medical imaging domains by only exploiting uncurated data such as CT scans and corresponding radiology reports. 

\section{Related Work}

\paragraph{Contrastive visual language learning.}

Contrastive learning \cite{oord2018representation, SimCLR} has evolved to be used in multi-modal data pipelines as popularised in CLIP \cite{CLIP}.
When applying CLIP, we are considering image and text modalities.
Recent work has extended the modalities to six \cite{imagebind}.
In essence, we are mapping multi-modal data to a unified latent space, where we calculate the similarities between different elements and learn models to map data across different modalities effectively.
When looking at how to calculate how similar different modalities are, we first look at the contrastive loss presented in Radford \etal \cite{CLIP}, which Sohn originally influenced \cite{n-loss} and then was further used in contrastive representation learning in Oord \etal \cite{inc-loss}.
An early attempt in CLIP radiology presented ConVIRT \cite{Convirt}, which maps chest X-rays to reports. 
CLIP's mainstream success has resulted in further applications to the medical domain \cite{pubmedclip}. 
CheXzero builds upon Radford \etal \cite{CLIP} by finetuning the model to the radiology domain, successfully achieving human-like results without explicit labels during training \cite{Chexzero}. 
Similarly, MedCLIP explores CLIP by applying alternative models to Open AI and CheXzero \cite{medclip}. 
In particular, they employ Swin transformer \cite{swin} and BioClinicalBERT \cite{bioclinicalbert} as their respective vision and text encoders.
They also perform ablation studies with a ResNet-50 \cite{resnet50} as the vision encoder, which provides the best results in zero-shot classification on COVID-19 and RSNA pneumonia X-ray datasets.
More recently, Zhang \etal built one of the most extensive medical image-pair datasets and obtained successful image-text/text-image retrieval results \cite{biomedclip}. 
One significant key difference is the increase in context length from 77 to 256.
Contrastive learning has also been successfully applied to lung CT images \cite{ct-cons2,ct-cons3}.
However, there is currently a lack of extensive literature exploring the application of this technique specifically to CT images and their corresponding reports.
Multi-modal learning provides opportunities to provide interesting insights into tasks such as zero-shot learning.

\paragraph{COVID-19.}

Mohit \etal \cite{9986406} use convolutional neural networks (CNNs) as encoders to build a computer-aided diagnosis system for COVID-19 to assist radiologists in early diagnosis.
Building upon this, transformers in supervised diagnosis, as seen in \cite{covid-vit}, can also be used, as shown in Dong \etal \cite{covid-trans}.
Moreso, this work demonstrates strong results on public datasets by extracting the relevant features from 3D volumes. 
We see contrastive learning used to build a severity system based on electronic health records (EHRs) \cite{WANYAN2021100389}.
Providing good results, structured data like this is often difficult to obtain, and we primarily focus more on unstructured data such as text.

\paragraph{Transformers.}
\label{text1}

Transformers have emerged as dominant models for natural language processing (NLP) and computer vision, largely due to their effective utilization of attention mechanisms \cite{attention1, attention2}.
In NLP settings, self-attention compares word embeddings to capture the relevance and importance of each word in the context \cite{attention}.
A popular choice, BERT \cite{BERT}, was recently fine-tuned on general and COVID-19-related radiology reports \cite{radBERT, covid-radbert}.
Influenced by text transformers, vision transformers have been argued to be more efficient in training than traditional CNNs \cite{vision}.
They divide images into fixed-size patches and utilize self-attention to capture the relationships between them.
This study considers three different vision transformers, ViT-B/16, ViT-B/32 \cite{vision} and Swin-Transformer \cite{swin}.
The architectures in these two transformers differ.
ViT operates on the entire image as a sequence of flattened patches, whereas Swin Transformer introduces a hierarchical structure of windows to process images. 
There have now been several implementations of vision transformers applied to medical images \cite{vitreview1, vitreview2}.


\section{Method}

\begin{figure}[t]
\begin{center}
\includegraphics[width=1.2\linewidth]{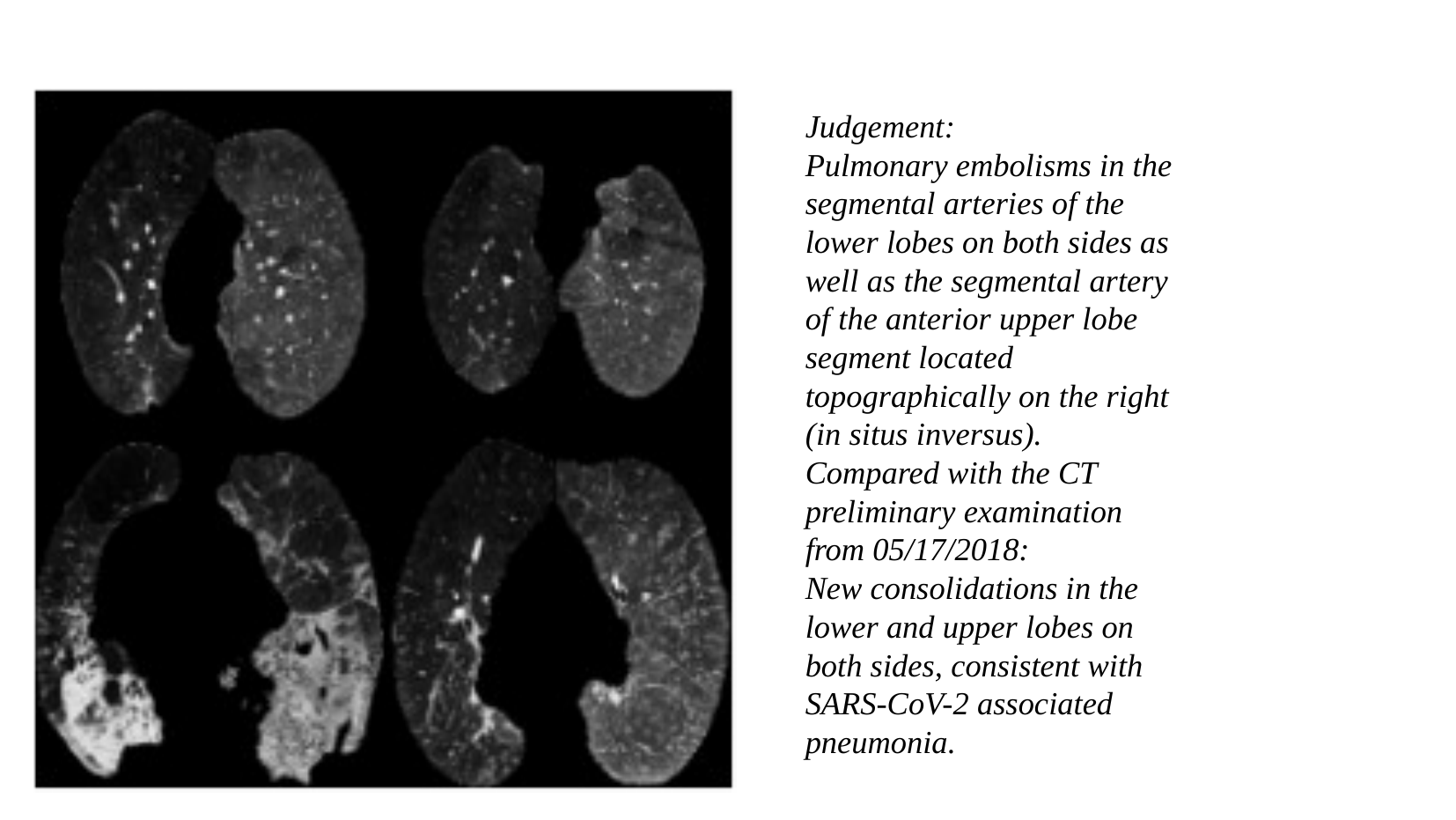}
\end{center}
   \caption{An example of the CT montage, four random slices taken from the CT image. The text is a subsection of the radiology report.}
 \label{fig:image-text-pair}
\end{figure}

\subsection{Data pre-processing} 
\label{data-process}

CT images are naturally large, given their multi-channel nature.
To decrease the memory overhead and filter out unwanted noise \cite{segmentation}, we employ a data pre-processing technique influenced by ILD diagnosis research \cite{Walsh}.
In particular, the pre-processing steps for each CT scan are as follows:

\begin{itemize}
    \item We first reduce the size along the axial dimension by 10\% on either end.
    CT scans contain redundant or low information at the beginning and end of the image.
    \item We then spatially crop the images to ensure there is no additional space and we are focusing on the lung tissue.
    \item The remaining CT is split into 4 blocks of similar dimensions. 
    \item A single slice is randomly selected from each block and concatenated to form a 4-image slice montage.
    \item Each 4-slice montage is resized to $224\times224$.
\end{itemize}

To speed up our data loading, we perform the aforementioned data pre-processing offline.
The pre-processing was performed 10 times for each CT scan.
Random slice selection can be considered a form of data augmentation.  
In Figure \ref{fig:image-text-pair}, we can see an example of the image-text pair.
For our text pre-processing, we translate the whole report from German to English using the Google Translate API in Python\footnote{The reports are originally in German, given that the hospitals are in the German speaking part of Switzerland.}. 
Then we slice the radiology report to focus only on the lung parenchyma.
Additionally, we filter the resulting text with filters taken from Clinical XLNet \cite{clinicalbert}.

\subsection{Encoders selection} \label{text2}

Our dataset is considered small compared to previous radiology datasets like CheXpert \cite{chexpert}, and MIMIC-CXR \cite{MIMIC-CXR}.
We compensate for the lack of data by building upon previous models specific to our task.
We consider established medical-based CLIP methods whilst also experimenting with extracting their vision encoder and finetuning with the RadBERT model.
The RadBERT model is pre-trained on 466 million tokens or 4.42M radiology reports \cite{radBERT}.
RadBERT was further fine-tuned in a COVID-19 investigation on 19,384 radiology reports this year in Chambon \etal \cite{covid-radbert}, making the model weights publicly available.
The prior representations of the text data learned in this model suit this study.

\begin{figure*}[t]
\begin{center}
\includegraphics[width=1\linewidth]{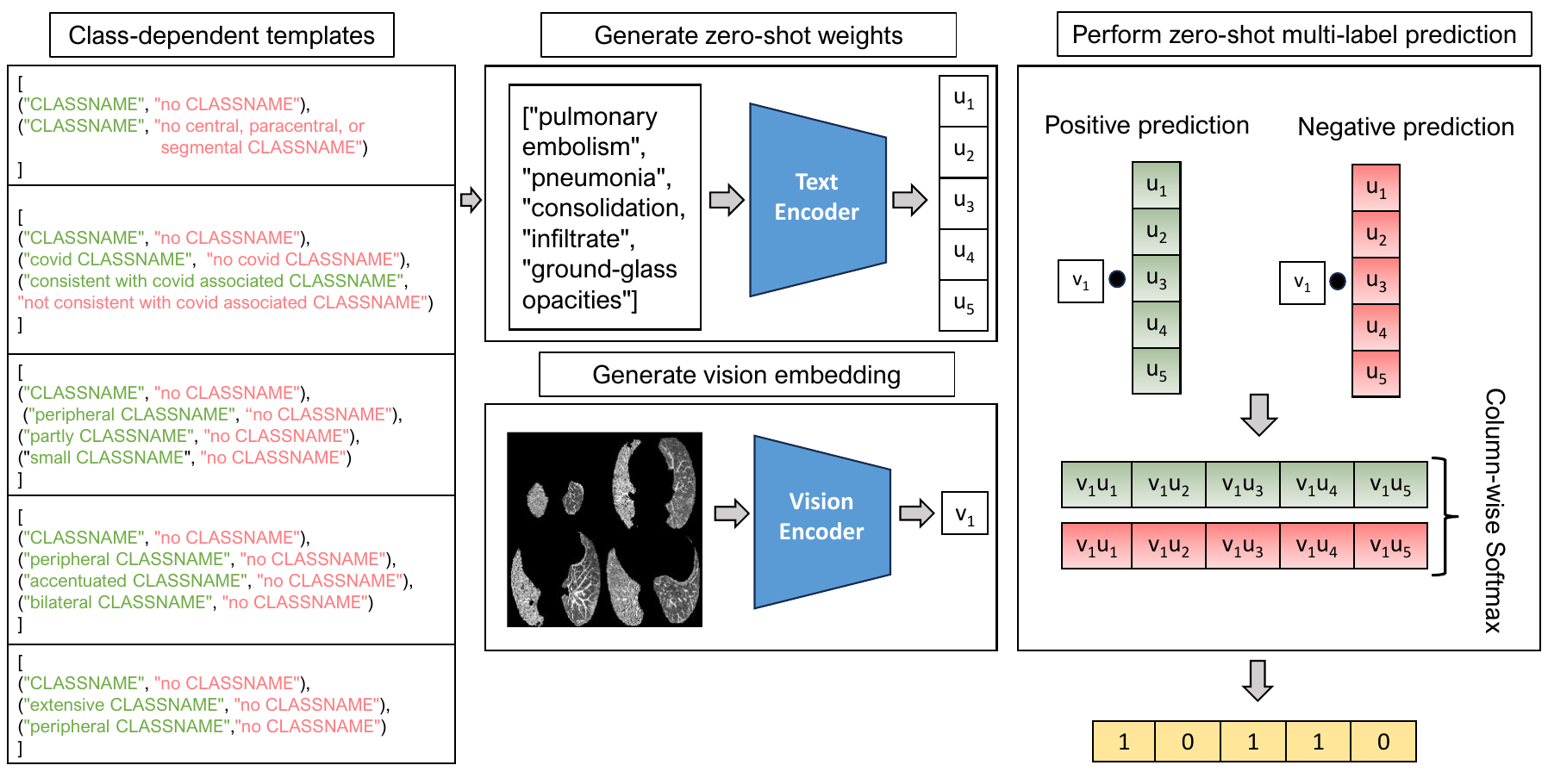}
\end{center}
   \caption{\textbf{Zero-shot pipeline.} In the left part, we see the five boxes which represent each class' list of template pairs. For a given class we iterate through its template pairs to generate the postive-negative zero-shot weights. Multiplying the respective CT montage embedding with both positive-negative zero-shot weights gives us a similarity. The last stage is calculating the softmax of these two predictions to result in our final prediction vector. }
 \label{fig:pipeline}
\end{figure*}

The choice of vision encoders is very large, but we are interested in those exposed to relevant data and tasks.
There is an observation of previous studies opting to use either vision transformers or standard CNNs such as ResNet-50 \cite{resnet50}. 
Precisely, we consider the following vision encoders: 

\begin{itemize}
    \item \textbf{ResNet-50 \cite{resnet50}.} Initialization with pre-trained weights on ImageNet \cite{russakovsky2015imagenet} is standard practice in computer vision.
    We set it up to compare to medical domain-trained models.
    
    \item \textbf{CheXzero \cite{Chexzero}.} We initialize the CLIP model \cite{CLIP} with publically available weights trained on the CheXpert and MIMIC-CXR datasets. The results in this paper are very promising in a multi-label task.
    We are only changing the image modality input. The text data is related to lung conditions specific to our radiology reports.

    \item \textbf{MedCLIP \cite{medclip}.} It considers both a ViT \cite{vision} and ResNet-50 \cite{resnet50} trained on their image-text dataset. The datasets picked are relevant to our task. Using alternative encoders, we are interested in the results compared to CheXzero.  
    
    \item \textbf{BiomedCLIP \cite{biomedclip}.} The BioMedCLIP set-up is attractive to us due to the size of the specific dataset it has been trained on. They have also increased their context length when training text encoders. 
    
    \item \textbf{COVID-ViT \cite{covid-vit}.} A custom ViT transformer used for COVID-19 diagnosis in CT slices. The model has learned to extract relevant features based on the pathological tissue in each CT slice.
    If correctly aligned, the vision encoder should successfully map the extracted image patterns to the relevant text features. 
\end{itemize}

When considering BiomedCLIP, MedCLIP, and CheXzero, we explore
the transfer without any adaptation, i.e., frozen encoders, the training of their vision and text encoders, and the extraction of only their vision encoder and training with the RadBERT text encoder. 
We also train the ResNet-50 and COVID-ViT with the RadBERT text encoder. 
When training with the RadBERT, we adapt each vision encoder to map the text output shape.

\subsection{Embeddings alignment} \label{loss}

Contrastive visual language training requires image and text embeddings to be mapped to the same latent space.
Closely following the forward pass in CLIP \cite{CLIP}, we align the output embeddings of the text and vision encoder by calculating the logits of each modality and passing these into separate cross-entropy loss functions.
During training, given a batch of $B$ input pairs $\left(\mathbf{x}_{v}, \mathbf{x}_{u}\right)$, we calculate their respective representation pairs $(\mathbf{v}, \mathbf{u})$ by feeding them into each respective encoder.
We use $(\mathbf{v_i}, \mathbf{u_i})$ to denote the $i$-th pair.
The first loss function is an image-to-text contrastive loss for the $i$-th pair:

$$
\ell_{i}^{(v \rightarrow u)}=-\log \frac{\exp \left(\left\langle\mathbf{v}_{i}, \mathbf{u}_{i}\right\rangle / \tau\right)}{\sum_{k=1}^{B} \exp \left(\left\langle\mathbf{v}_{i}, \mathbf{u}_{k}\right\rangle / \tau\right)},
$$

while similarly, the text-to-image loss:

$$
\ell_{i}^{(u \rightarrow v)}=-\log \frac{\exp \left(\left\langle\mathbf{u}_{i}, \mathbf{v}_{i}\right\rangle / \tau\right)}{\sum_{k=1}^{B} \exp \left(\left\langle\mathbf{u}_{i}, \mathbf{v}_{k}\right\rangle / \tau\right)}
$$

where $\left\langle\mathbf{v}_{i}, \mathbf{u}_{i}\right\rangle$ represents the cosine similarity and $\tau \in \mathbb{R}^{+}$ represents a temperature parameter.
The temperature parameter controls the range of the logits in the stated losses and the strength of penalties on hard negative samples \cite{temperature}.
To calculate the overall loss, we calculate the average of the losses.
Precisely, we add them together and divide them by the number of modalities, i.e., two. 

\subsection{Zero-shot multi-label classification}

\begin{figure*}[t]
    \centering
    \resizebox{0.48\linewidth}{!}{\includegraphics{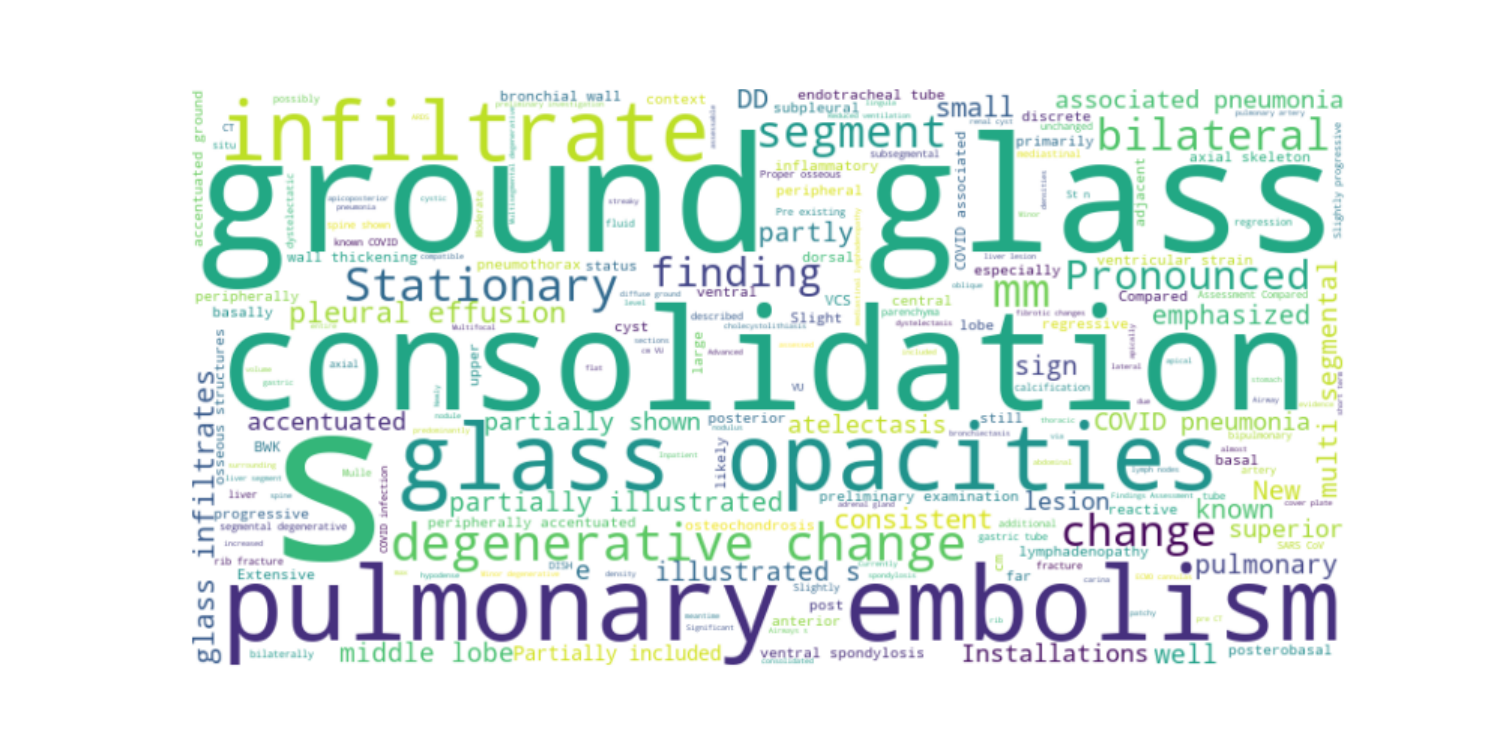}}
    \resizebox{0.48\linewidth}{!}{\includegraphics{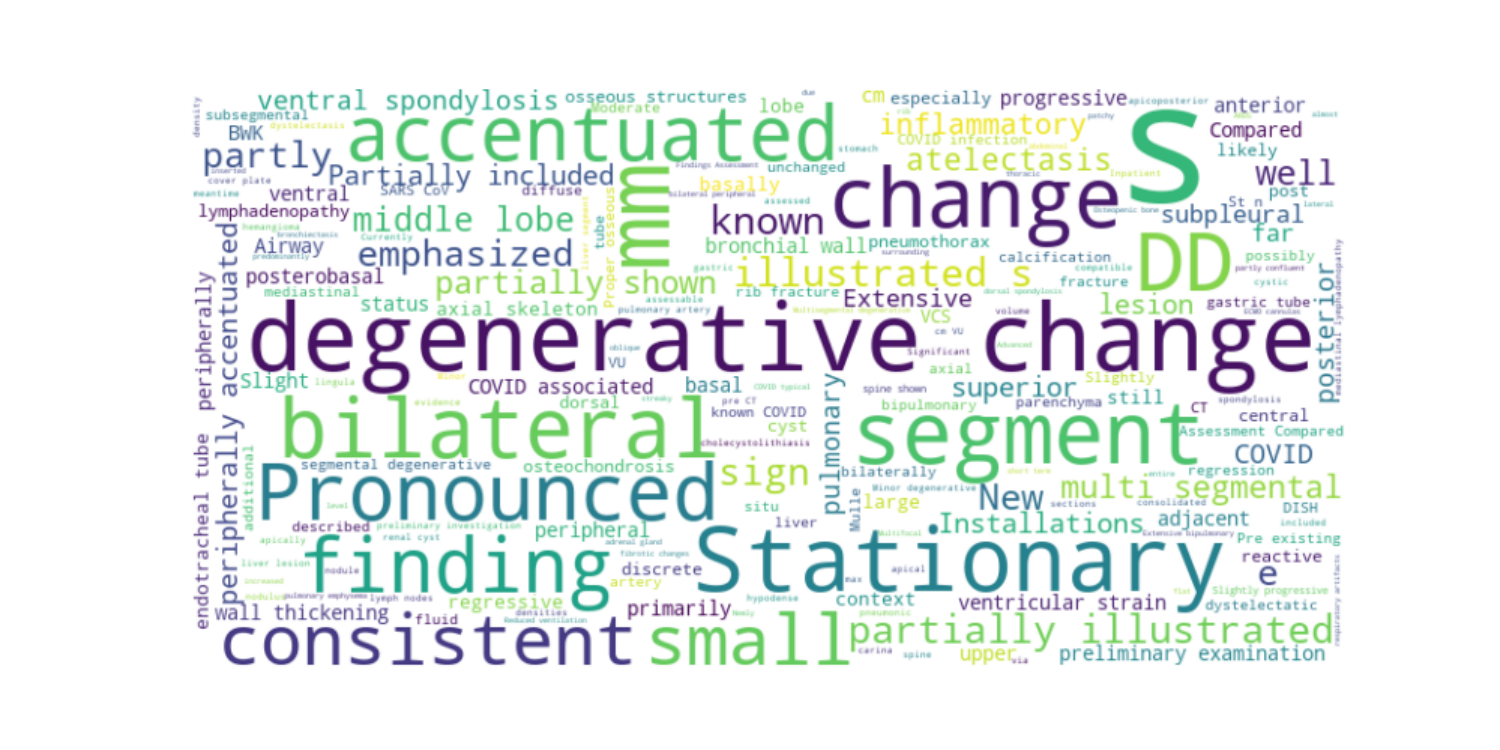}}

   \caption{\textbf{Word Clouds.} Word cloud generated to assist in class decisions (left).
   Word cloud with class names removed to assist in template selection (right).
   }
 \label{fig:word-cloud}
\end{figure*}

Zero-shot methods have gained traction recently, popularised by seminal papers such as Socher \etal \cite{zero-shot} and Radford \etal \cite{CLIP}.
The difficulty in evaluating our models lies within the embedding space of our text data.
Previous works have utilized shorter, curated text data.
This allows the design of simple prompts that match the training distribution, e.g., ``A picture of a CAT".
However, our report data comes directly from radiologists and reflects a realistic medical setting.
To overcome this challenge, we employ prompt-based engineering, as seen in the latest approaches \cite{Lu_2023_CVPR, Chexzero, CLIP}.
In collaboration with human experts, we analyze the radiology reports to create sensible classes for the dataset.
We first create a word cloud (Figure \ref{fig:word-cloud}, left) on the text data to visualize the most common words.
We settle on the following classes: pulmonary embolism, pneumonia, consolidation, infiltrates and ground glass opacities. 
Confirming these are sensible and useful classes for radiologists, the human experts label the testing samples for us.
Pulmonary embolism is the least common class, diagnosed in approximately one in ten patients. 
The prevalence of the other classes ranges from 65\% to 80\%.
These classes lay unstructured in the text data we are training on and remain hidden from the vision encoders as in traditional zero-shot learning \cite{Wang2019ASO}.
For a given class, our prompt is a positive-negative template paired with the word CLASSNAME, which is replaced by the class.
The choice of templates required manual analysis of the reports to estimate what prompts would be a good choice.
To justify our template choice, we removed the class names from the text data and generated a word cloud visible in Figure \ref{fig:word-cloud} (right).
Using this and with the manual reading of the reports, we identify prompts which occur with our class names; for example, the phrase ``bilateral infiltrates" creates our prompt ``bilateral CLASSNAME" for the class infiltrates.

Following Tiu \etal \cite{Chexzero}, to map the models' prediction to probabilities, we use a softmax layer for each template pair.
Instead of using the same template pairs for each class, we propose each class has a list of its template pairs.
This is visualized in Figure \ref{fig:pipeline}.
More in detail: 

\begin{itemize}
    \item We iterate over each class. In each class, we iterate over the template pairs.
    For a given template, we substitute the class into the template (e.g., no pulmonary embolism) and pass it into the text encoder. 
    We normalize the embeddings and concatenate the results to get our zero-shot weights. 
    We are left with a 5D vector which is our zero-shot weights for each class.
    \item We pass the respective CT montage into the vision encoder and normalize the embeddings.
    \item We calculate the cosine similarity of the resulting embeddings by multiplying the zero-shot weights and the vision embeddings.
    \item We estimate the class by calculating the softmax of the positive-negative predictions.
\end{itemize}

We do not scale by the temperature parameter $\tau$ in the zero-shot evaluation.
We do not include this as we are relying on the representations learned by our encoders.
The parameter is no longer needed to control the range of the logits.
As a CT montage can have more than one label, we solve a multi-label zero-shot problem by performing binary classification for each class.


 \begin{table}[!ht]
    \centering
    \begin{tabular}{cc}
    \toprule
        Variable &  All patients \\ \midrule
        Age (years $\pm$ SD) & 61.4 $\pm$ 14.2\\ 
         Male / Female & 276 / 84  \\ 
         BMI ($\frac{Kg}{m^2}$ $\pm$ SD) & 27.9 $\pm$ 8.5 \\
         Data split (Training / test) & 368 / 92 \\
    \bottomrule
    \\
    \end{tabular}
    \caption{Descripive statistics of the patient cohort.}
    \label{table_patients}
\end{table}

\section{Experimental Settings}

 \begin{table*}[t]
    \centering

    \resizebox{\linewidth}{!}{
    \begin{tabular}{ccccccc}
    \toprule
    Model & Encoders & Fine-tuned & Context length & Macro Avg. F1 ($\uparrow$) & HL ($\downarrow$) & Sub. Acc. ($\uparrow$) \\ \midrule
    CheXzero \cite{Chexzero} & ViT-B/32 $\lvert$ GPT2& \ding{55} & 77 & 0.43 & 0.54 & 0.00 \\
    CheXzero \cite{Chexzero} & ViT-B/32 $\lvert$ GPT2 & \ding{51} & 77 & 0.71 & 0.29 & 0.23 \\ \midrule
    MedCLIP \cite{medclip} & Swin-T $\lvert$ BCB & \ding{55} & 77 & 0.62 & 0.45 & 0.04 \\
    MedCLIP \cite{medclip} &  Swin-T $\lvert$ BCB & \ding{51} & 77 & 0.51 & 0.48 & 0.12 \\ \midrule
    BioMedCLIP \cite{biomedclip} & ViT-B/16 $\lvert$ PMB & \ding{55} & 256 & 0.33 & 0.49 & 0.06 \\
    BioMedCLIP \cite{biomedclip} & ViT-B/16 $\lvert$ PMB & \ding{51} & 256 & 0.61 & 0.44 & 0.09 \\ \bottomrule
    \\
    \end{tabular}
    }
    \caption{\textbf{Frozen vs fine-tuned encoders}. Comparison between the frozen and fine-tuned models.
    HL stands for the Hamming loss, and Sub. Acc. for subset accuracy. 
    PMB, BCB respectively, denote PubMedBERT and BioClinicalBERT.}
    \label{baselines}
\end{table*}

\subsection{Dataset}
After the ethics approval, the study collected data from University Hospital Zurich in Switzerland, resulting in over 460 image-text pairs from individual patients taken during the COVID-19 pandemic. 
A small summary of the patient cohort can be seen in \cref{table_patients}.
CT scans were taken on either Siemens SOMATOM Definition AS, SOMATOM Definition Flash, or SOMATOM Force  machines.
We only consider thorax CT images taken in a single session.
After resampling the CT images and preparing the montages (\cref{data-process}), we are left with 13,240 montages and 460 reports. 
We split the dataset 80:20 based on patient ID, so montages from the same CT are not be seen in training and evaluation. 
We trained on 11,010 montages along with their respective reports. 
When evaluating our test set, we chose a single montage-report pair for each patient and performed the zero-shot evaluation.

\subsection{Training details}

In all training scenarios, we set a maximum of 100 epochs, a batch size of 100, and use the AdamW optimizer \cite{adamW} with betas of (0.9, 0.98).
When fine-tuning existing methods (CheXzero, MedCLIP, BioMedCLIP), we adapt the hyperparameters to achieve the best results. Specifically, we set the learning rate to [5e-6, 5e-5, 5e-5] and the weight decay to [1e-4, 1e-4, 1e-3], respectively. For training RadBERT and alternative vision encoders, we fix the learning rate at 5e-5 and the weight decay at 1e-3.

We implemented offline image data preparations to accelerate data loading. 
At the bottom of the image processing pipeline described in \cref{data-process}, we convert montages to PIL images and tensors.
During training, we use a single NVIDIA RTX A6000 GPU to train the encoders.


\subsection{Zero-shot evaluation}

To assess the zero-shot capabilities, we consider calculating the macro average F1 score, Hamming loss, and the subset accuracy. 
The F1 score is defined as 
$$
F1 = \frac{2 * Precision * Recall}{Precision + Recall}
$$
Where precision and recall are respectively defined as $\frac{TP}{TP + FP}$ and  $\frac{TP}{TP + FN}$.
The macro average F1 gives us the average of the F1 over all classes. 

Hamming loss is often considered in multi-label evaluation. 
The Hamming loss measures the fraction of instances where a model's predictions do not equal the true labels. 
It is obtained by dividing the number of incorrect predictions by the total number of instances and classes.
$$HL = \frac{1}{N L} \sum_{l=1}^L\sum_{i=1}^N Y_{i,l} \oplus X_{i,l},$$
$N$ is the total number of data samples and $L$ is the total number of classes.
$\oplus$ is Exlusive-OR, $X_{i,l}$ ($Y_{i,l}$) stands for boolean that the $i$-th index (prediction) is equal to the $l$-th label.

Lastly, the subset accuracy measures how accurately every sample is predicted. 
For instance, if our target array is [0, 1, 1, 0, 0], it is considered correct only if all elements of the vector are predicted accurately.
The prediction [0, 1, 1, 1, 0] is hence considered wrong. 
This makes the subset accuracy the most challenging metric to fulfill.


 \begin{table}[!ht]
    \centering
    \resizebox{\linewidth}{!}{
    \begin{tabular}{cccccccccc}
    \toprule
    Model  & Templates & Macro Avg. F1 & HL & Sub. Acc. \\ \midrule
    \multirow{2}{4em}{CheXzero} & CI & 0.55 & 0.47 & 0.05\\
    & CD & 0.71 & 0.29 & 0.23 \\ \midrule
    \multirow{2}{4em}{RN50} & CI & 0.46 & 0.51 & 0.12 \\
    & CD & 0.50 & 0.45 & 0.21 \\ \midrule
    \multirow{2}{4em}{MedCLIP}& CI & 0.57 & 0.44 & 0.13 \\ 
    & CD & 0.58 & 0.42 & 0.15 \\ \bottomrule
    \\
    \end{tabular}
    }
    \caption{\textbf{Class-independent (CI) vs. class-dependent (CD) templates}. CheXzero is the ViT-B/32 plus GPT2. The ResNet-50 (RN50) is pre-trained with ImageNet. MedCLIP is the Swin transformer. The RN50 and MedCLIP are coupled with the RadBERT text encoder with a context length of 100 and 200, respectively.
    All models are fine-tuned on our dataset.}
    \label{class}
\end{table}

\section{Results}

 \begin{table*}[!ht]
    \centering
    \resizebox{\linewidth}{!}{
    \renewcommand{\arraystretch}{0.5}
    \begin{tabular}{cccccccccc}
    \toprule
    Model & Vision Encoder & Context length & Truncation side & Macro Avg. F1 ($\uparrow$) & HL ($\downarrow$) & Sub. Acc. ($\uparrow$)\\ \midrule
    \multirow{4}{4em}{CheXzero\\\cite{Chexzero}} & \multirow{4}{4em}{ViT-B/32} & 100 & $\leftarrow$ & 0.53 & 0.52 & 0.08 \\
    & & 100 & $\rightarrow$ & 0.59 & 0.45 & 0.14 \\
    & & 200 & $\leftarrow$ & 0.54 & 0.48 & 0.09 \\
    & & 200 & $\rightarrow$ & 0.57 & 0.46 & 0.13 \\ \midrule
    \multirow{4}{4em}{MedCLIP \cite{medclip}} & \multirow{4}{4em}{RN50} & 100 & $\leftarrow$ & 0.53 & 0.49 & 0.12 \\
    &  & 100 & $\rightarrow$ & 0.55 & 0.48 & 0.13 \\
    &  & 200 & $\leftarrow$ & 0.46 & 0.56 & 0.08 \\
    &  & 200 & $\rightarrow$ & 0.54 & 0.51 & 0.09 \\ \midrule
    \multirow{4}{4em}{MedCLIP \cite{medclip}} & \multirow{4}{4em}{Swin-T} & 100 & $\leftarrow$ & 0.5 & 0.51 & 0.09 \\
    &  & 100 & $\rightarrow$ & 0.59 & 0.46 & 0.09 \\
    &  & 200 & $\leftarrow$ & 0.58 & 0.42 & 0.15 \\
    &  & 200 & $\rightarrow$ & 0.53 & 0.53 & 0.09 \\ \midrule
    \multirow{4}{4em}{BioMedCLIP\\\cite{biomedclip}} & \multirow{4}{4em}{ViT-B/16} & 100 & $\leftarrow$ & 0.48 & 0.53 & 0.10 \\
    & & 100 & $\rightarrow$ & 0.42 & 0.59 & 0.10 \\
    & & 200 & $\leftarrow$ & 0.50 & 0.53 & 0.05 \\
    & & 200 & $\rightarrow$ & 0.51 & 0.51 & 0.10 \\ \midrule
    \multirow{4}{4em}{CovidViT\\\cite{covid-vit}} & \multirow{4}{4em}{ViT} & 100 & $\leftarrow$ & 0.51 & 0.49 & 0.10 \\
    &  & 100 & $\rightarrow$ & 0.55 & 0.45 & 0.10 \\
    &  & 200 & $\leftarrow$ & 0.49 & 0.52 & 0.05 \\
    &  & 200 & $\rightarrow$ & 0.54 & 0.49 & 0.10 \\ \midrule
    \multirow{4}{4em}{--} & \multirow{4}{4em}{RN50\\IN-1k} & 100 & $\leftarrow$ & 0.51 & 0.54 & 0.06 \\
    &  & 100 & $\rightarrow$ & 0.50 & 0.49 & 0.21 \\
    &  & 200 & $\leftarrow$ & 0.53 & 0.51 & 0.12 \\
    &  & 200 & $\rightarrow$ & 0.45 & 0.54 & 0.08 \\ \midrule
    \end{tabular}
    }
    \caption{\textbf{Combining encoders}. Comparison between models with fixed text encoder (RadBERT) and multiple vision encoders.
    The left and right arrows, respectively, mean truncation from the end and beginning of reports.
    HL stands for the Hamming loss, and Sub. Acc. for subset accuracy.
    RN50 IN-1k refers to the ResNet-50 initialized from ImageNet pre-trained weights.
    PMB, BCB respectively, denote PubMedBERT and BioClinicalBERT.}
    \label{combined}
\end{table*}

In this section, we empirically investigate the best training solution for reliably mapping a relatively small set of CT images to corresponding radiology reports and consequently performing zero-shot predictions.
As anticipated, we consider three approaches: 1) applying frozen baseline methods, 2) finetuning baseline methods, and 3) training an alternative text and vision encoder.
We also study the impact of class-dependent and class-independent templates.
We perform additional ablations considering the truncation side and context length.

\subsection{Baselines and finetuning}

For a comprehensive comparison, we consider three baselines that provide publically available weights applying contrastive visual language models in the medical domain. 
CheXzero \cite{Chexzero}, MedCLIP \cite{medclip}, and BioMedCLIP \cite{biomedclip} are all available and are relevant to our task.

\textbf{Frozen encoders.} As baseline methods, we load the pre-trained models and apply straight-away the zero-shot evaluation on our test set without any modifications.
All evaluated baseline models feature transformers.
The results for the frozen encoders are shown in \cref{baselines}.
We see the subset accuracies are extremely low, from a minimum of 0\% to a maximum of 6\%.
However, considering none of the models has been exposed to CT montages before, we see the effect of large datasets does not compensate for the shift to our domain, and the models need specific finetuning.

\textbf{Fine-tuned encoders.} Next, we finetune each baseline on our small dataset and again perform the evaluation.
As expected, we observe an increase in the exact match performance (\cref{baselines}). 
MedCLIP and BioMedCLIP respectively gain 8 and 3 percent points.
Notably, CheXzero has adapted extremely well to our dataset, increasing the subset accuracy from 0\% to 23\%. 
The other metrics are also the best we observe in this study, i.e., Hamming loss of 0.29 and macro average F1 score of 0.71.
This finding suggests that previous pre-training on large chest X-ray datasets and their corresponding reports helps achieve better results in this fine-grained task.
Unexpectedly, the macro average F1 and Hamming loss for MedCLIP decreases.
We suspect that the concurrent decrease of these two metrics and the increase in subset accuracy imply a recognition improvement for a subset of the classes and a deterioration in others.
We finally see a big increase in the macro average F1 score for BioMedCLIP but not so much in the other metrics.

\textbf{Class-independent vs. class-dependent templates.}
In contrast to previous studies, each zero-shot class has its list of template pairs. 
This enables us better target vision features to specific prompts for a class. 
For example, the positive prompt for pneumonia was often "consistent with", whereas this prompt would not make sense for a different class like a pulmonary embolism.
As shown in \cref{class}, CheXzero improves drastically in all three metrics, in particular with a subset accuracy increase of 18\%, when applying class-dependent prompts.
All previous results, shown in \cref{baselines}, are obtained with class-dependent templates. 

\subsection{Combining pre-trained models.}

Investigating the combination of pre-trained models trained specifically on COVID radiology reports and chest image datasets is required to identify the best setups.
All text encoders use the COVID-finetuned RadBERT, which has prior knowledge of radiological terms such as ``pulmonary embolism" and ``ground glass opacities". 
We also select the existing pre-trained CLIP models in the medical domain and extract their vision encoders. 
These vision encoders have all been exposed to chest X-rays, and some have even been exposed to single CT slices.
Furthermore, we analyze a vision transformer used in COVID diagnosis and a pre-trained ResNet-50 on ImageNet. 
The difficulty of the task is acknowledged in the results achieved, receiving the best subset accuracy of 21\% (\cref{combined}). 
Interestingly, the latter is obtained with RN50, which has not been pre-trained on medical datasets.

We run further ablations considering text and multiple vision encoders.
We experiment with class-dependent vs. class-independent templates, changing the context length and truncating from different sides. 

\textbf{Class-independent vs. class-dependent templates.} 
\cref{class} also shows positive trends for the top-two combinations in \cref{combined}, those trained with RadBERT and an alternative vision encoder. 
With regards to subset accuracy, the ResNet-50 increases by 9\% and MedCLIP marginally improves by 2\%. 
All three metrics improve when applying class-dependent prompts. 

\textbf{Context length and truncation.} Due to our uncurated radiology reports being longer than standard image-text pairs, we test varying context lengths equal to 100 or 200. 
Previous work in contrastive visual language learning deals with shorter text, so we explore increasing this to capture more information. 
Furthermore, we test whether truncating the tokens from the left or the right affects the performance.
At the beginning of the reports, we can see the knowledge describing the lungs.
At the end of the report, we see shorter statements that match our classes.
We test whether more valuable representations can be learned from the beginning or end.

Our results show the best subset accuracy result with the shorter context length and truncating from the right. 
The best Hamming loss was recorded with a longer context length, truncated from the left, and still records a good subset accuracy.  
The average subset accuracy for context length 100 is 9.3\%, and for 200, it is 9.41\%. 
Equally, for truncation on the left, it is 7.8\%, and for the right, it is 9.6\%.
We conclude that context length did not alter the results massively, but the truncation side worked better when truncating from the right.

\section{Conclusions}

In our paper, we investigated the development of a zero-shot tool that assists radiologists in detecting pulmonary embolisms and identifying intricate lung details, including ground glass opacities and consolidations, automatically.

To meet this goal, we collected uncurated COVID-19 CT scans and corresponding reports from a university hospital to ensure our study faces real-world scenarios.
Secondly, we run tests based on image-text pretraining and fine-tuning for multi-label zero-shot classification.
A clear challenge represented the variability of the text data and the consequent framing of the zero-shot targets.
We partially addressed such an issue using a class-dependent zero-shot template scheme and pre-trained vision-text medical models.
In parallel, we are in the process of curating data from three additional hospitals. 
This data will enable us to further develop and apply the described methods to longitudinal data, specifically for the prognosis of long COVID-19. 
In future work, we would like to see such techniques applied to 3D volumes to solve tasks such as disease prognostication and progression as discussed recently in review articles.\cite{prog1, prog2}.

We hope our work inspires future research to meaningfully use data collected throughout the pandemic and improve the automatic identification of fine-grained lung pathological patterns. 

\section{Acknowledgements}

This work was supported in part by the Emergency Department and the Department of Diagnostic, Interventional, and Pediatric Radiology of Inselspital Bern and in part by Campus Stiftung Lindenhof Bern (SLB).
{\small
\bibliographystyle{ieee_fullname}
\bibliography{egbib}

\begin{thebibliography}{10}\itemsep=-1pt

\bibitem{covid-trans}
Dong A, Liu J, Zhang G, Wei Z, Zhai Y, and Lv G.
\newblock Momentum contrast transformer for covid-19 diagnosis with knowledge
  distillation.
\newblock {\em Pattern Recognit.}, 2023.

\bibitem{bioclinicalbert}
Emily Alsentzer, John~R. Murphy, Willie Boag, Wei-Hung Weng, Di Jin, Tristan
  Naumann, and Matthew B.~A. McDermott.
\newblock Publicly available clinical bert embeddings.
\newblock {\em arXiv:1904.03323}, 2019.

\bibitem{attention1}
Dzmitry Bahdanau, Kyunghyun Cho, and Yoshua Bengio.
\newblock Neural machine translation by jointly learning to align and
  translate.
\newblock 2014.

\bibitem{covid-radbert}
Pierre Chambon, Tessa~S. Cook, and Curtis~P. Langlotz.
\newblock Improved fine‑tuning of in‑domain transformer model for inferring
  covid‑19 presence in multi‑institutional radiology reports.
\newblock 2019.

\bibitem{SimCLR}
Ting Chen, Simon Kornblith, Mohammad Norouzi, and Geoffrey Hinton.
\newblock A simple framework for contrastive learning of visual
  representations.
\newblock 2020.

\bibitem{CVPR2022}
Elijah Cole, Xuan Yang, Kimberly Wilber, Oisin~Mac Aodha, and Serge Belongie.
\newblock When does contrastive visual representation learning work?
\newblock 2022.

\bibitem{BERT}
Jacob Devlin, Ming-Wei Chang, Kenton Lee, and Kristina Toutanova.
\newblock Bert: Pre-training of deep bidirectional transformers for language
  understanding.
\newblock 2018.

\bibitem{vision}
Alexey Dosovitskiy, Lucas Beyer, Alexander Kolesnikov, Dirk Weissenborn,
  Xiaohua Zhai, Thomas Unterthiner, Mostafa Dehghani, Matthias Minderer, Georg
  Heigold, Sylvain Gelly, Jakob Uszkoreit, and Neil Houlsby.
\newblock An image is worth 16x16 words: Transformers for image recognition at
  scale.
\newblock 2020.

\bibitem{prog1}
Dack E, Christe A, Fontanellaz M, Brigato L, Heverhagen JT, Peters AA, Huber
  AT, Hoppe H, Mougiakakou S, and Ebner L.
\newblock Artificial intelligence and interstitial lung disease: Diagnosis and
  prognosis.
\newblock {\em Invest Radiol.}, 2023.

\bibitem{CVPR2021}
Linus Ericsson, Henry Gouk, and Timothy~M. Hospedales.
\newblock How well do self-supervised models transfer?
\newblock 2021.

\bibitem{pubmedclip}
Sedigheh Eslami, Gerard de Melo, and Christoph Meinel.
\newblock Does clip benefit visual question answering in the medical domain as
  much as it does in the general domain?
\newblock 2021.

\bibitem{covid-vit}
Xiaohong Gao, Yu Qian, and Alice Gao.
\newblock Covid-vit: Classification of covid-19 from ct chest images based on
  vision transformer models.
\newblock 2021.

\bibitem{imagebind}
Rohit Girdhar, Alaaeldin El-Nouby, Zhuang Liu, Mannat Singh, Kalyan~Vasudev
  Alwala, Armand Joulin, and Ishan Misra.
\newblock Imagebind: One embedding space to bind them all.
\newblock 2023.

\bibitem{prog2}
Barnes H, Humphries SM, George PM, Assayag D, Glaspole I, Mackintosh JA, Corte
  TJ, Glassberg M, Johannson KA, Calandriello L, Felder F, Wells A, and Walsh
  S.
\newblock Machine learning in radiology: the new frontier in interstitial lung
  diseases.
\newblock {\em Lancet Digit Health.}, 2023.

\bibitem{resnet50}
Kaiming He, Xiangyu Zhang, Shaoqing Ren, and Jian Sun.
\newblock Deep residual learning for image recognition.
\newblock 2015.

\bibitem{vitreview1}
Emerald~U. Henry, Onyeka Emebob, and Conrad~Asotie Omonhinmin.
\newblock Vision transformers in medical imaging: A review.
\newblock 2022.

\bibitem{segmentation}
Johannes Hofmanninger, Florian Prayer, Jeanny Pan, Sebastian Rohrich, Helmut
  Prosch, and Georg Langs.
\newblock Automatic lung segmentation in routine imaging is primarily a data
  diversity problem, not a methodology problem.
\newblock 2020.

\bibitem{clinicalbert}
Kexin Huang, Abhishek Singh, Sitong Chen, Edward~T. Moseley, Chih ying Deng,
  Naomi George, and Charlotta Lindvall.
\newblock Clinical xlnet: Modeling sequential clinical notes and predicting
  prolonged mechanical ventilation.
\newblock 2019.

\bibitem{self-supervised-review}
Shih-Cheng Huang, Anuj Pareek, Malte Jensen, Matthew~P. Lungren, Serena Yeung,
  and Akshay~S. Chaudhari.
\newblock Self-supervised learning for medical image classification: a
  systematic review and implementation guidelines.
\newblock 2023.

\bibitem{chexpert}
Jeremy Irvin, Pranav Rajpurkar, Michael Ko, Yifan Yu, Silviana Ciurea-Ilcus,
  Chris Chute, Henrik Marklund, Behzad Haghgoo, Robyn Ball, Katie Shpanskaya,
  Jayne Seekins, David~A. Mong, Safwan~S. Halabi, Jesse~K. Sandberg, Ricky
  Jones, David~B. Larson, Curtis~P. Langlotz, Bhavik~N. Patel, Matthew~P.
  Lungren, and Andrew~Y. Ng.
\newblock Chexpert: A large chest radiograph dataset with uncertainty labels
  and expert comparison.
\newblock 2019.

\bibitem{covid-review2}
Kaur J and Kaur P.
\newblock Outbreak covid-19 in medical image processing using deep learning: A
  state-of-the- art review.
\newblock 2020.

\bibitem{Lijing}
Li Jing, Pascal Vincent, Yann LeCun, and Yuandong Tian.
\newblock Understanding dimensional collapse in contrastive self-supervised
  learning.
\newblock 2022.

\bibitem{MIMIC-CXR}
Alistair E.~W. Johnson, Tom~J. Pollard, Seth~J. Berkowitz, Nathaniel~R.
  Greenbaum, Matthew~P. Lungren, Chih ying Deng, Roger~G. Mark, and Steven
  Horng.
\newblock Mimic-cxr, a de-identified publicly available database of chest
  radiographs with free-text reports.
\newblock 2019.

\bibitem{ct-cons3}
Sota Kato, Masahiro Oda, Kensaku Mori, Akinobu Shimizu, Yoshito Otake, Masahiro
  Hashimoto, Toshiaki Akashi, and Kazuhiro Hotta.
\newblock Classification and visual explanation for covid-19 pneumonia from ct
  images using triple learning.
\newblock 2022.

\bibitem{swin}
Ze Liu, Yutong Lin, Yue Cao, Han Hu, Yixuan Wei, Zheng Zhang, Stephen Lin, and
  Baining Guo.
\newblock Swin transformer: Hierarchical vision transformer using shifted
  windows.
\newblock {\em arXiv:2103.14030}, 2021.

\bibitem{adamW}
Ilya Loshchilov and Frank Hutter.
\newblock Decoupled weight decay regularization.
\newblock {\em arXiv:1711.05101}, 2017.

\bibitem{Lu_2023_CVPR}
Lu, Ming Y., Chen, Bowen, Zhang, Andrew, Williamson, Drew~F. K., Chen, Richard
  J., Ding, Tong, Le, Long Phi, Chuang, Yung-Sung, Mahmood, and Faisal.
\newblock Visual language pretrained multiple instance zero-shot transfer for
  histopathology images.
\newblock In {\em Proceedings of the IEEE/CVF Conference on Computer Vision and
  Pattern Recognition (CVPR)}, pages 19764--19775, June 2023.

\bibitem{attention2}
Minh-Thang Luong, Hieu Pham, and Christopher~D. Manning.
\newblock Effective approaches to attention-based neural machine translation.
\newblock 2015.

\bibitem{9986406}
Mohit, Kumar, Gupta, Rajeev, Kumar, and Basant.
\newblock Self-supervised contrastive learning for covid-19 classification from
  computed tomography images.
\newblock In {\em 2022 IEEE 9th Uttar Pradesh Section International Conference
  on Electrical, Electronics and Computer Engineering (UPCON)}, pages 1--5,
  2022.

\bibitem{inc-loss}
Oord, A. v. d., Li, Y., , Vinyals, and O. R.
\newblock Representation learn- ing with contrastive predictive coding.
\newblock 2018.

\bibitem{oord2018representation}
Oord, Aaron van den, Li, Yazhe, Vinyals, and Oriol.
\newblock Representation learning with contrastive predictive coding.
\newblock {\em arXiv preprint arXiv:1807.03748}, 2018.

\bibitem{vitreview2}
Arshi Parvaiz, Muhammad~Anwaar Khalid, Rukhsana Zafar, Huma Ameer, Muhammad
  Ali, and Muhammad~Moazam Fraz.
\newblock Vision transformers in medical computer vision—a contemplative
  retrospection.
\newblock 2022.

\bibitem{CLIP}
Alec Radford, Jong~Wook Kim, Chris Hallacy, Aditya Ramesh, Gabriel
  Gohand~Sandhini Agarwal, Girish Sastry, Amanda Askell, Pamela Mishkin, Jack
  Clarkand~Gretchen Krueger, and Ilya Sutskever.
\newblock Learning transferable visual models from natural language
  supervision.
\newblock 2021.

\bibitem{russakovsky2015imagenet}
Russakovsky, Olga, Deng, Jia, Su, Hao, Krause, Jonathan, Satheesh, Sanjeev,
  Maand Sean, Huang, Zhiheng, Karpathy, Andrej, Khosla, Aditya, Bernstein,
  Michael, et~al.
\newblock Imagenet large scale visual recognition challenge.
\newblock {\em International Journal of Computer Vision (IJCV)}, 2015.

\bibitem{Walsh}
Walsh SLF, Calandriello L, Silva M, and Sverzellati N.
\newblock Deep learning for classifying fibrotic lung disease on
  high-resolution computed tomography: a case-cohort study.
\newblock 2018.

\bibitem{zero-shot}
Richard Socher, Milind Ganjoo, Hamsa Sridhar, Osbert Bastani, Christopher~D.
  Manning, and Andrew~Y. Ng.
\newblock Zero-shot learning through cross-modal transfer.
\newblock 2013.

\bibitem{n-loss}
Kihyuk Sohn.
\newblock Improved deep metric learning with multi-class n-pair loss objective.
\newblock 2016.

\bibitem{ct-cons2}
Xie T, Wei Y, Xu L, Li Q, Che F, Xu Q, Cheng X, Liu M, Yang M, Wang X, Zhang F,
  Song B, and Liu M.
\newblock Self-supervised contrastive learning using ct images for pd-1/pd-l1
  expression prediction in hepatocellular carcinoma.
\newblock 2023.

\bibitem{Chexzero}
Ekin Tiu, Ellie Talius, Pujan Patel, Curtis~P. Langlotz, Andrew~Y. Ng, and
  Pranav Rajpurkar.
\newblock Expert-level detection of pathologies from unannotated chest x-ray
  images via self-supervised learning.
\newblock 2020.

\bibitem{attention}
Ashish Vaswani, Noam Shazeer, Niki Parmar, Jakob Uszkoreit, Llion Jones,
  Aidan~N. Gomez, Lukasz Kaiser, and Illia Polosukhin.
\newblock Attention is all you need.
\newblock 2017.

\bibitem{wang2020covid}
Wang, Linda, Lin, Zhong Qiu, Wong, and Alexander.
\newblock Covid-net: A tailored deep convolutional neural network design for
  detection of covid-19 cases from chest x-ray images.
\newblock {\em Scientific reports}, 2020.

\bibitem{wang2017chestx}
Wang, Xiaosong, Peng, Yifan, Lu, Le, Lu, Zhiyong, Bagheri, Mohammadhadi,
  Summers, and Ronald M.
\newblock Chestx-ray8: Hospital-scale chest x-ray database and benchmarks on
  weakly-supervised classification and localization of common thorax diseases.
\newblock In {\em Proceedings of the IEEE conference on computer vision and
  pattern recognition}, 2017.

\bibitem{temperature}
Feng Wang and Huaping Liu.
\newblock Understanding the behaviour of contrastive loss.
\newblock 2021.

\bibitem{Wang2019ASO}
Wei Wang, Vincent~Wenchen Zheng, Han Yu, and Chunyan Miao.
\newblock A survey of zero-shot learning.
\newblock {\em ACM Transactions on Intelligent Systems and Technology (TIST)},
  10:1 -- 37, 2019.

\bibitem{medclip}
Zifeng Wang, Zhenbang Wu, Dinesh Agarwal, and Jimeng Sun.
\newblock Medclip: Contrastive learning from unpaired medical images and text.
\newblock 2022.

\bibitem{WANYAN2021100389}
Tingyi Wanyan, Hossein Honarvar, Suraj~K. Jaladanki, Chengxi Zang, Nidhi Naik,
  Sulaiman Somani, Jessica~K. {De Freitas}, Ishan Paranjpe, Akhil Vaid, Jing
  Zhang, Riccardo Miotto, Zhangyang Wang, Girish~N. Nadkarni, Marinka Zitnik,
  Ariful Azad, Fei Wang, Ying Ding, and Benjamin~S. Glicksberg.
\newblock Contrastive learning improves critical event prediction in covid-19
  patients.
\newblock {\em Patterns}, 2(12):100389, 2021.

\bibitem{covid-review1}
Laure Wynants, Ben~Van Calster, and Gary~S Collins.
\newblock Prediction models for diagnosis and prognosis of covid-19: systematic
  review and critical appraisal.
\newblock 2020.

\bibitem{radBERT}
An Yan, Julian McAuley, Xing Lu, Jiang Du, Eric~Y. Chang, Amilcare Gentili, and
  Chun-Nan Hsu.
\newblock Radbert: Adapting transformer-based language models to radiology.
\newblock 2022.

\bibitem{biomedclip}
Sheng Zhang, Yanbo Xu, Naoto Usuyama, Jaspreet Bagga, Robert Tinn, Sam Preston,
  Rajesh Rao, Mu Wei, Naveen Valluri, Cliff Wong, Matthew~P. Lungren, Tristan
  Naumann, and Hoifung Poon.
\newblock Large-scale domain-specific pretraining for biomedical
  vision-language processing.
\newblock 2023.

\bibitem{Convirt}
Yuhao Zhang, Hang Jiang, Yasuhide Miura, Christopher~D. Manning, and Curtis~P.
  Langlotz.
\newblock Contrastive learning of medical visual representations from paired
  images and text.
\newblock 2020.

\end{thebibliography}
}

\end{document}